\documentclass[11pt,psfig,leqno,amsmath,amssymb]{article}
\usepackage{graphics}
\usepackage{graphicx}
\usepackage{bm}
\usepackage{epsf}
\newcommand{\f}{\frac}

\newcommand{\half}{\frac{1}{2}}
\newcommand{\beq}{\begin{equation}}
\newcommand{\eeq}{\end{equation}}
\newcommand{\bem}{\begin{displaymath}}
\newcommand{\eem}{\end{displaymath}}
\newcommand{\bey}{\begin{eqnarray}}
\newcommand{\eey}{\end{eqnarray}}

\newcommand{\D}{\Delta}

\newcommand{\PP}{\mbox{Pr}}

\def\l{\lambda}
\def\f{\frac}

\def\bm#1{\mbox{\boldmath{$#1$}}}
\def\D#1#2{{\frac{d #1}{d #2}}}

\begin{document}

\title{\bf
Generalized Gas Dynamic Equations for Microflows}
\author{
Kun Xu\footnote{email: makxu@ust.hk} \\
Mathematics Department  \\
Hong Kong University of Science and Technology  \\
Clear Water Bay, Kowloon, Hong Kong \\
Zhaoli Guo\footnote{email:zlguo@mail.hust.edu.cn} \\
National Laboratory of Coal Combustion \\
 Huazhong University of
Science and Technology, Wuhan 430074, China }

\date{}

\maketitle

\bigskip\bigskip
 \centerline{\sl { }}



\section*{Abstract}

In an early approach, we proposed a kinetic model with multiple
translational temperature [K. Xu, H. Liu and J. Jiang, Phys. Fluids
{\bf 19}, 016101 (2007)], to simulate non-equilibrium flows. In this
paper, instead of using three temperatures in $x-$, $y-$, and
$z$-directions, we are going to further define the translational
temperature as a second-order symmetric tensor.  Based on a multiple
stage BGK-type collision model and the Chapman-Enskog expansion, the
corresponding macroscopic gas dynamics equations in
three-dimensional space will be derived. The zeroth-order expansion
gives the 10 moment closure equations of Levermore [C.D. Levermore,
J. Stat. Phys {\bf 83}, pp.1021 (1996)]. To the 1st-order expansion,
the derived gas dynamic equations can be considered as a
regularization of Levermore's 10 moments equations. The new  gas
dynamic equations have the same structure as the Navier-Stokes
equations, but the stress strain relationship in the Navier-Stokes
equations is replaced by an algebraic equation with temperature
differences. At the same time, the heat flux, which is absent in
Levermore's 10 moment closure, is recovered. As a result, both  the
viscous and the heat conduction terms are unified under a single
anisotropic temperature concept. In the continuum flow regime, the
new gas dynamic equations automatically recover the standard
Navier-Stokes equations. The current gas dynamic equations are
natural extension of the Navier-Stokes equations to the near
continuum flow regime and can be used for microflow computations. 
Two examples, the force-driven Poiseuille flow and the Couette flow
in the transition flow regime, are used to validate the model. Both
analytical and numerical results are encouraging.

\section{Introduction}

The transport phenomena, i.e., mass, heat, and momentum transfer, in
the different flow regime is of a great scientific and practical
interest. The classification of  various flow regimes is based on
the dimensionless parameter, i.e., the Knudsen number, which is a
measure of the degree of rarefaction of the medium. The Knudsen
number $Kn$ is defined as the ratio of the mean free path to a
characteristic length scale of the system. In the continuum flow
regime where $Kn < 0.001$, the Navier-Stokes equations with linear
relations between stress and strain and the Fourier's law for heat
conduction are adequate to model the fluid behavior. For flows in
the continuum-transition regime ($0.1 < Kn < 1$), the Navier-Stokes
(NS) equations are known to be inadequate. This regime is important
for many practical engineering problems. Hence, there is a strong
desire and requirement for accurate models which give reliable
solutions with low computational costs.


One of the alternative approach to simulate the non-equilibrium flow
is those based on the moment closures. Grad's 13 moment equations
are one of the most important ones, which provide the time evolution
of the non-equilibrium quantities, such as the stress and the heat
flux \cite{grad}. However, due to its hyperbolic nature, these
equations lead to a well known sub-shocks problem inside a shock
layer as the Mach number is larger than a critical value. In order
to improve the validity of the 13 moment equations, based on the
Chapman-Enskog expansion, Struchtrup and Torrilhon introduced terms
of super-Burnett order to the balance of pressure deviator and heat
flux vector in the moment equations, and got the regularized  $13$
moment (R13) equations which have much better performance in the
non-equilibrium flow regime \cite{struchtrup}. Another well-known
moment system is Levermore's 10 moment closure, which follows his
hierarchy of non-perturbative moment closures with many desirable
mathematical properties \cite{levermore}. For example, these
equations don't suffer from the closure-breakdown deficiencies, and
they always give physically realizable solution due to non-negative
gas distribution function. However, the 10 moment Gaussian closure
has no heat flux even though it is proved that Navier-Stokes viscous
terms can be recovered in the continuum flow regime. In an effort to
extend the Gaussian closure to include higher-order effects, Groth
et. al. formulated perturbative variants to the original moment
closure with new extended fluid dynamic model \cite{groth}. And the
most well studied of these closures is a 35-moment closure.
Recently, McDonald and Groth took a Chapman-Enskog-type expansion
about either the moment equations or the kinetic equation, and
introduced the heat flux into Levermore's 10 moment closures and
obtained extended fluid dynamic equations for non-equilibrium flow
simulation \cite{mcdonald}. The new system present improved results
in the transition flow regime where the  heat transfer has a
significant effect.

In recent years, we have concentrated on the development of
numerical schemes for the near continuum flow simulation.  In order
to capture the non-equilibrium physics in the transitional flow
regime, we have extended the gas-kinetic Navier-Stokes flow solver
with the following developments \cite{xu01}. First, a closed
solution of the gas distribution function up to the NS order has
been used to derive a generalized particle collision time,
subsequently to obtain the extended viscosity and heat conduction
coefficients \cite{xu-pof}. Later, in order to describe the
non-equilibrium flow related to the molecular rotational and
vibrational degree of freedom, a multiple time relaxation kinetic
scheme has been introduced for the shock structure calculations
\cite{xu06,cai}. Recently, the gas-kinetic scheme has been further
extended to study the multiple translational temperature
non-equilibrium \cite{xu07}. The schemes developed in the above
study give reasonable results in the transitional flow regime, such
as the capturing of shock structure at different Mach numbers and
flow phenomena which cannot be described properly by the NS
equations. In the above simulations, the underlying physical model
is a generalized BGK (GBGK) model \cite{xu07}, where the multiple
stage relaxation processes have been considered.

In the  gas-kinetic schemes, the solutions are obtained without
knowing the explicit macroscopic governing equations. In this paper,
following the numerical procedure we are going to fill up the gap
and derive the underlying macroscopic governing equations for the
monatomic gas.  In an early approach \cite{xu07}, we introduced the
multiple translational temperature into the kinetic model, where the
energy exchanges between $x$-, $y$-, and $z$-directions are modeled
through the particle collision. Based on the above kinetic model, in
one dimensional space the generalized NS equations are derived,
where the viscous term in the NS equations is replaced by the
temperature relaxation term. In this paper, we will further develop
such a model and regard the temperature as a second order tensor.
Since the gas flow may settle to an equilibrium state through
multiple stages \cite{xu08}, one of the reasonable assumption is to
use a Gaussian distribution with multiple temperature as a middle
state. Physically, this state corresponds to a gas with different
temperature in different directions. Therefore, the thermal energy
or particle random motion is represented through a symmetric
temperature tensor $T_{ij}$. Due to historical reason, the
temperature is defined as a thermodynamical variable, where there is
no bulk fluid velocity variations. So, the temperature becomes a
scalar concept. However, in the transition flow regime the gas has
large bulk velocity variation, and there is no enough particle
collisions to equalize the random particle motion. In order to
construct gas dynamic equations, it is justified to extend the
temperature from a scalar concept to a tensor. The commutable
property of the random particle velocity determines the temperature
to be a symmetric tensor. Actually, this kind of non-equilibrium gas
property  has been routinely extracted from the DSMC solutions. So,
based on the physical model with the Gaussian distribution as a
middle state between the real gas distribution function $f$ and the
equilibrium Maxwellian $f^{eq}$ and the strategy used in the
construction of the kinetic scheme, we are going to derive the
corresponding macroscopic governing equations. Surprisingly, the
obtained gas dynamic  equations become regularization of Levermore's
10 moment Gaussian closure, where additional viscous and heat
conduction terms are obtained. The structure of gas dynamic
equations are almost identical to the Navier-Stokes equations, but
the constitutive relationship of the NS one, i.e., $\sigma_{ij} =
-\rho R T^{eq} \delta_{ij} + \mu (\partial_i U_j +
\partial_j U_i - \frac{2}{3} \partial_k U_k \delta_{ij} ) $, is
replaced by the new one $\sigma_{ij} = -\rho R T_{ij} + \rho R
(T^{eq}_{ij} - T_{ij}  ) $. At the same time, the heat flux depends
on the gradient of the temperature $T_{ij}$.

This paper is organized in the following. Section 2 is about the
introduction of kinetic equation and the generalized particle
collision model. At the same time, the 10 moment closure and the
generalized gas dynamic equations will be presented. Section 3 is
about the applications of the new gas dynamic equations to two flow
problems in the near continuum regime. The last section is the
conclusion.

\section{Generalized Gas Dynamic Equations }

In this section, we first review the particle collision model,
introduce the Gaussian closure, and derive the new gas dynamic
equations. A monatomic gas will be considered in this paper.

\subsection{Two stage gas-kinetic collision model }

The gas-kinetic Bhatnagar-Gross-Krook (BGK) model has the form
\cite{BGK},
\beq
\partial_t f + u_i \partial_i f  = ({f^{eq} - f})/{\tau},
 \label{eq:bgk}
 \eeq
where the particle distribution function $f$ is a function of time
$t$, spatial location $x_i$, and particle velocity $u_i$. The left
hand side of the above equation represents the free streaming of
molecules in space, and the right side denotes the simplified
collision term of the Boltzmann equation.  In the BGK model, the
collision operator involves a single  relaxation time $\tau$ for a
non-equilibrium state to evolve to an equilibrium one $f^{eq}$,
which is an isotropic Gaussian
$$f^{eq} = \frac{\rho}{(2\pi R T^{eq} )^{3/2}} \exp [ -\frac{(u_i-U_i)(u_i-U_i)}{2
RT^{eq} } ],$$ where $\rho$ is the density, $T^{eq}$ is the
equilibrium temperature, and $U_i$ is the averaged macroscopic fluid
velocity. Traditionally, based on the above BGK model, the
Navier-Stokes and higher-order equations, such as Burnett and
Super-Burnett, can be derived \cite{chapman,ohwada}. Unfortunately,
these higher-order equations have intrinsic physical and
mathematical problems in the transitional flow regime. In general,
the BGK collision term is valid only for flows close to the thermal
equilibrium one. In order to extend the capacity of the BGK model to
the non-equilibrium flow regime, we can re-write the collision term
of the BGK model into two physical sub-processes,
\beq
 \partial_t f  + u_i \partial_i f = ({g - f})/{\tau}+({f^{eq} - g})/{\tau},
 \label{eq:bgk-T1}
 \eeq
where $g$ is a middle state between $f$ and $f^{eq}$, see figure 1.
The DSMC solutions show that the randomness of particle motion
depends on the spatial direction. So, a nature assumption about the
middle state is a state with multiple temperature. In the above
equation, the term  $(f^{eq} - g)/\tau$ has no direct connection
with $f$, therefore, we can consider it as a source term in the
above generalized BGK (GBGK) model,
\beq
 \partial_t f + u_i \partial_i f = ({g - f})/{\tau}+ Q,
 \label{eq:bgk-T}
 \eeq
where $Q=(f^{eq}-g)/\tau$ for the monatomic gas.

In an early approach, we have assumed that the middle state $g$ has
individual temperature in $x-$, $y-$ and $z-$directions, and it has
the following form in a two dimensional case,
\beq
 g = \rho \left( \f{\l_x}{\pi} \right)^{1/2} \left( \f{\l_y}{\pi}\right)^{1/2}
\left( \f{\l_z}{\pi} \right)^{1/2} \exp \left[-\l_x (u-U)^2 -\l_y
(v-V)^2 -\l_z w^2 \right]. \label{eq:gg}
 \eeq
Here $\l_x = 1/(2 RT_x), \l_y =1/(2RT_y)$, and $\l_z = 1/(2RT_z)$
are related to the translational temperature $T_x,T_y$, and $T_z$ in
$x$, $y$, and $z$ directions. Here $R$ is the gas constant.

In order to solve the above kinetic equation numerically to
determine the time evolution of the macroscopic physical quantities
$(\rho, U, V, T_x, T_y,T_z)$, we proposed the following scheme.
First, we expand the gas distribution function around the multiple
temperature state $g$, such as $f = g -\tau (g_t + u_i \partial_i g
)$, from which the numerical fluxes across each cell interface are
evaluated. Second, the source term $Q$ is integrated in a time step
inside each computational cell and is regarded as a source term. The
physical effect of the source term is to equalize the temperature in
different directions. The above processes are composed of the
relaxations from the non-equilibrium state $f$ to a multiple
temperature state $g$, then $g$ converges to an equal temperature
Maxwellian. These two processes may have different relaxation time
scales, and the specific formulation of $g$ depends on the flow
problems. According to the above model, we have derived the
hydrodynamic equations in one dimensional space \cite{xu07}, where
the stress-strain relationship in the Navier-Stokes equations is
replaced by the temperature relaxation. In the continuum flow
regime, where the middle state is close to the Maxwellian, the
standard Navier-Stokes equations can be recovered. The numerical
tests presented in \cite{xu07} verified the validity of the above
model. In this paper, we are going to use multiple temperature
Gaussian as the middle state, and according to the similar procedure
to derive general gas dynamic equations in three dimensional space.

To approximate the Boltzmann particle collision term by  multiple
BGK-type sub-processes have been investigated before by many
authors, such as Callaway \cite{callaway}, Gorban and Karlin
\cite{gorban}, and Levermore \cite{levermore}. For the DSMC method,
to include the rotation and vibration modes has been done through
the BGK-type relaxation model as well. As realized in
\cite{levermore}, for Maxwellian molecules, the use of the multiple
BGK-type collision term is correct even if the flow is far away from
the equilibrium. This suggests that the generalized BGK operator may
be a legitimate approximation to the collision term for use in the
near continuum flow regime.

\subsection{Generalized gas dynamic equations: zeroth order}

For the rarefied flow simulation, a generalized middle state $g$
between $f$ and $f^{eq}$ can be a Gaussian distribution. The
Gaussian distribution appears to have been derived in the early work
by Maxwell \cite{maxwell}, and then re-discovered by many
researchers, such as Holway \cite{holway} and Levermore
\cite{levermore}. In this paper, we directly define the gas
temperature as a tensor instead of a scalar. The traditional
temperature concept is coming from thermodynamics, where the local
bulk fluid velocity deviation is absent. However, for the transport
equations, the temperature basically represents the degree of
randomness of particle motion. In order to dynamically equalize the
temperature, there should have enough particle collision. In the
near continuum flow regime, the number of particle collision inside
a microscale device, is limited. It is most likely that the
particles will keep the non-isotropic particle randomness.

 One of the middle state we can use between $f$ and $f^{eq}$ is the
Gaussian distribution,
$$g = \frac{\rho}{\sqrt{\det(2\pi R T_{ij} )}}
\exp (- \half (u_i-U_i) (RT_{ij})^{-1} (u_j-U_j) ) ,$$ where
$T_{ij}$ is the positive definite temperature matrix, which is
related to the thermal energy of the particle motion $\rho R T_{ij}
= \int (u_i -U_i)(u_j -U_j)g d u $. Due to the commutable property
between particle randomness velocities, such as $(u_i-U_i)(u_j-U_j)
= (u_j-U_j)(u_i-U_i) $, $T_{ij}$ must be a symmetric tensor. In the
Levermore's approach \cite{levermore}, he developed  a
non-perturbative method, where the gas distribution function $f$ is
assumed to be equal to $g$. Based on the kinetic equation
(\ref{eq:bgk-T}) with the assumption $f=g$, i.e.,
$$ \partial_t g + u_i \partial_i g = Q,$$
taking the moments
$${\psi} = (1, u_i,  u_i u_j )^T ,$$
on the above equation gives the following Gaussian closure,
\beq \partial_t \rho + \partial_k (\rho U_k )  =0 , \label{eq:con}
 \eeq
\beq \partial_t (\rho U_i ) + \partial_k [\rho (U_i U_k + R T_{ik})]
=0 , \label{eq:mom}
 \eeq
\beq \partial_t [\rho (U_i U_j + RT_{ij} )] + \partial_k [ \rho (U_i
U_j U_k + R U_k T_{ij} + R U_i T_{jk} + R U_j T_{ki} ) ]
=\frac{1}{\tau}  \rho R (T^{eq}\delta_{ij} - T_{ij} ) .
\label{eq:ene}
 \eeq
These equations are actually the zeroth-order Chapman-Enskog
expansion, i.e., $f=g$, for the generalized BGK model.  In the above
equations, the source term on the right hand side is coming from the
term $Q$ in the generalized BGK model. The equilibrium temperature
$T^{eq}$ is defined by
$$T^{eq} =\frac{1}{3} \mbox{Tr} (T_{ij}).$$
If the state $g$ is equal to the Maxwellian, i.e., $g=f^{eq}$, the
above equations reduce to the Euler equations. Based on the above
equations, Levermore and Morokoff shows  that if the initial data of
$T_{ij}$ is symmetric positive definite, then it remains so
\cite{levermore2}.

%

Even without heat conduction terms, Levermore showed that the above
equations recover the Navier-Stokes viscous terms in the continuum
flow regime \cite{levermore}. For the above system
(\ref{eq:con})-(\ref{eq:ene}), the left hand side equations have a
complete real eigenvalues and eigenvectors \cite{brown,hittinger},
and the system is strictly hyperbolic. The temperature difference in
different direction, such as the $T_{ij}$, basically shows that the
sound speed depends on the spatial direction, the so-called
anisotropic wave propagation due to the non-isotropic gas property.

 In order to further introduce  heat flux into the above
10 moment closure, McDonald and Groth recently took a
Chapman-Enskog-type expansion about the Gaussian moment equations
\cite{mcdonald}. The heat conduction is added for the thermal
temperature equation,
\beq
\partial_t T_{ij} + \partial_k (U_k T_{ij} ) + T_{ik} \partial_k U_j
+ T_{jk} \partial_k U_i +\partial_k Q_{ijk} = \frac{1}{\tau} (T^{eq}
\delta_{ij} - T_{ij} ), \label{eq:tem}
 \eeq
where the heat flux $Q$ has the form
$$Q_{ijk} = -\frac{\tau}{\mbox{Pr}}
[T_{kl} \partial_l T_{ij} + T_{jl} \partial_l T_{ik} + T_{il}
\partial_l T_{jk} ].
$$

\subsection{Generalized gas dynamic equations: first order}

The generalized BGK model includes two relaxation process. One is
from $f$ to the Gaussian $g$, and the other is from $g$ to an
equilibrium state $f^{eq}$. In the last section, the distribution
function $f$ is assumed to be equal to $g$, and the 10 moment
closure equations are derived. However, as presented in figure 1,
the real distribution function should be different from $g$, and the
process of relaxation from $f$ to $g$ has to be considered. In the
past years, we have developed gas-kinetic schemes based on the
generalized  BGK model, where a gas distribution function $f$ around
$g$ has been constructed and used to evaluate the numerical fluxes
in a finite volume scheme \cite{xu07}. The schemes present
reasonable numerical solutions in the near continuum flow regime,
such as the micro-channel flow computation \cite{xu08}.  In the
following, we are going to derive the corresponding macroscopic
governing equations underlying the gas-kinetic scheme. The method
used here can be regarded as the Chapman-Enskog expansion or
iterative expansion \cite{ohwada}, which are equivalent to each
other.

The solution $f$ around the Gaussian $g$ is constructed using the
iterative expansion to the 1st-order order \cite{ohwada},
\beq
 f = g - \tau  (\partial_t g+  u_i \partial_i g )+\tau Q,
\label{eq:chap} \eeq
where in the kinetic scheme $\partial_t g$ is determined using the
compatibility condition
\beq \int \psi (\partial_t g + u_i \partial_i g ) du = \int \psi Q
du , \label{eq:m} \eeq
 which is exactly the 10 moment closure
(\ref{eq:con})-(\ref{eq:ene}).
 Substituting the distribution function $f$ in (\ref{eq:chap}) into the BGK
model (\ref{eq:bgk-T}), the equation becomes
\beq \partial_t g + u_i  \partial_i  g = \tau ( \partial_t^2  g + 2
u_i
\partial_t \partial_i g + u_i u_j \partial_i \partial_j g ) + Q -\tau (\partial_t Q + u_i \partial_i Q).
\label{eq:1st} \eeq
Taking the moments $\psi$ to the above equation and using
Eq.(\ref{eq:m}) to express the time derivative in terms of the
spatial derivative, we can get the following macroscopic equations,
\beq \partial_t \rho + \partial_k (\rho U_k )  =0 , \label{eq:con-r}
 \eeq
\beq \partial_t (\rho U_i ) + \partial_k [\rho (U_i U_k + R T_{ik}
)] =
\partial_k [\rho R (T^{eq} \delta_{ki} - T_{ki} ) ]  , \label{eq:mom-r}
 \eeq
\bey  & & \partial_t  [ \rho (U_i U_j + RT_{ij} )] + \partial_k [
\rho (U_i U_j U_k + R U_k T_{ij} + R U_i T_{jk} + R U_j T_{ki} ) ]
\nonumber
\\
&  & = \frac{2}{\tau}  \rho R (T^{eq} \delta_{ij} - T_{ij} ) \nonumber \\
& & \hskip 0.3cm +
\partial_k \{\rho R [U_k (T^{eq}\delta_{ij} - T_{ij} ) + U_i (T^{eq}
\delta_{jk} - T_{jk} ) + U_j (T^{eq} \delta_{ki} - T_{ki} )  ] \}\nonumber \\
& & \hskip 0.2cm +  \partial_k \{  \tau \rho R^2 \frac{1}{\mbox{Pr}}
(T_{kl}
\partial_l T_{ij} + T_{il}
\partial_l T_{jk} + T_{jl}
\partial_l T_{ki} ) \} .
\label{eq:ene-r}
 \eey
The above equations have been written in a similar way as the
Navier-Stokes equations. The differences between the above equations
and the 10 moment closure are the additional terms appeared on the
right hand sides of the momentum and energy equations. It is
interesting to see that the corresponding heat conduction term
derived above has the same form as that obtained by McDonald and
Groth, even though they are obtained through different
considerations \cite{mcdonald}. Based on the BGK-type collision
model, a unit Prandtl number is obtained for the heat conduction
term. However, since we believe that up to the NS order the
structure of the gas dynamic equations will not be changed due to
the BGK collision term or the exact Boltzmann collision model, we
add the Prandtl number in the above corresponding heat flux term.
Since the viscosity and heat conduction coefficients are the
concepts for the continuum flow, the particle collision time $\tau$
in the above equations is defined according to the result in the
continuum regime,
$$\tau = \mu / (\rho R T^{eq} ) ,$$
where $\mu$ is the dynamical viscosity coefficient. Certainly, in
the rarefied flow regime, the corresponding viscosity coefficient
has to be modified \cite{xu-pof}. Also, attention should be paid on
the relaxation term in the energy equation, where additional $2$
appears.

The above equations can be re-arranged to get the time evolution
equation for the thermal energy $\rho  R T_{ij}$,
\bey  & & \partial_t [ \rho T_{ij} ] + \partial_k [ \rho U_k T_{ij}]
\nonumber
  \\ & & =
 \frac{2}{\tau} \rho  (T^{eq}\delta_{ki} - T_{ij} ) \nonumber \\
 & & \hskip 0.5cm -\rho [ T_{kj}
\partial_k U_i + T_{ki} \partial_k U_j ] \hskip 8.1cm (a) \nonumber \\
& & \hskip 0.5cm +
\partial_k [\rho  U_k (T^{eq}\delta_{ij} - T_{ij} ) ]
+ \rho  (T^{eq} \delta_{jk} - T_{jk} ) \partial_k U_i + \rho (T^{eq}
\delta_{ki} - T_{ki} )\partial_k U_j \hskip 0.5cm (b)
\nonumber \\
& & \hskip 0.6cm + \partial_k \{ \tau \rho \frac{R}{\mbox{Pr}}
(T_{kl}
\partial_l T_{ij} + T_{il}
\partial_l T_{jk} + T_{jl}
\partial_l T_{ki} )  \}  \hskip 4.7cm (c) .
 \label{eq:ene-t}
\eey
In comparison with the Navier-Stokes equations, the right hand side
of the equation (\ref{eq:ene-t}) has a clear physical meaning. Here
(a). the forces multiplied by fluid deformation, which is the
heating and cooling of the fluid by compression or expansion and
this term represents a reversible process. (b). viscous dissipation,
which is responsible for heat generation, and it is always positive
and produces internal energy. This is an irreversible process. (c).
heat conduction term, which is also irreversible.

 The relaxation
parameter $\tau$ controls the distance between the non-equilibrium
state $f$ and the equilibrium one $f^{eq}$. The current method for
the derivation of the gas dynamic equations can be also used to
other system, such as these with a more complicated non-equilibrium
middle state $g$, such as the distributions with $14$ or $26$
moments. However, a distribution function $g$ with higher-order
terms may correspond to macroscopic governing equations without
clear physical meanings for the higher-order terms.

The generalized gas dynamic equations
(\ref{eq:con-r})-(\ref{eq:ene-r}) have the same left hand side as
the 10 moment closure equations (\ref{eq:con})-(\ref{eq:ene})
\cite{levermore2}. And the fluxes on the left have a complete
eigenvalues and eigenvectors \cite{brown,hittinger,mcdonald2}. In
other words, the left hand side is the hyperbolic part. For the
right hand side, besides the heat flux as recently derived by
McDonald and Groth \cite{mcdonald}, additional dissipative terms,
i.e., $\rho R (T^{eq} \delta_{ij} - T_{ij} )$ in the momentum
equations, and $\rho R [U_k (T^{eq}\delta_{ij} - T_{ij} ) + U_i
(T^{eq} \delta_{jk} - T_{jk} ) + U_j (T^{eq} \delta_{ki} - T_{ki} )]
$ in the thermal energy equation, have been obtained. Furthermore,
in the present model the relaxation time appearing in the first term
on the right hand side of Eq.(\ref{eq:ene-r}) is $\tau/2$, while the
corresponding term in the model by McDonald and Groth is $\tau$. In
our two-stage collision model. If we compare the above equations
(\ref{eq:con-r})-(\ref{eq:ene-r}) with the Navier-Stokes equations,
where the energy equation can be also separated  in a
direction-by-direction componentwise form, we can immediately
realize that the corresponding viscous term is given by
$$ \sigma_{ij}^{'}  = \rho R (T^{eq} \delta_{ij} - T_{ij} ) .$$
Therefore, the constitutive relationship for the new gas dynamic
equations is
\beq  \sigma_{ij}  = -\rho R T_{ij} + \rho R (T^{eq} \delta_{ij} -
T_{ij}) . \label{eq:stress-n} \eeq
In the following, we are going to show that the new constitutive
relationship can recover the standard NS formulation in the
continuum flow regime.

In the continuum flow regime, the Gaussian distribution $g$ will
come to the same state as the equilibrium one $f^{eq}$, see figure
1. In this case, the 1st-order expansion of $f$ presented in this
section will be expanded basically around $f^{eq}$. To the leading
order of small $\tau$, Eq.(\ref{eq:ene-r}) gives the temperature
deviation,
\bey & & \rho R (T^{eq}\delta_{ij} - T_{ij} ) \approx \nonumber \\
& & (\tau/2) \{
\partial_t  [ \rho (U_i U_j + RT_{ij} )] + \partial_k [
\rho (U_i U_j U_k + R U_k T_{ij} + R U_i T_{jk} + R U_j T_{ki} ) ]
\}.
 \eey
In the above equation, when applying the equilibrium condition
$T_{ii} = T^{eq}$ and $T_{ij} = T^{eq} \delta_{ij} $, and using the
Euler equations to replace the temporal derivative by spatial
derivative, we can get
\bey & & \rho R (T^{eq}\delta_{ij} - T_{ij} ) \nonumber \\
& &
\approx (\tau /2) \rho R T^{eq}
 [ \partial_i U_j + \partial_j  U_i - \frac{2}{3} \partial_k U_k
 \delta_{ij}] = (\mu /2) [ \partial_i U_j + \partial_j  U_i - \frac{2}{3} \partial_k U_k
 \delta_{ij}].
 \label{eq:ns-s}
 \eey
Therefore, the constitutive relationship (\ref{eq:stress-n}) becomes
\bey  \sigma_{ij}  &=& -\rho R T^{eq} \delta_{ij} + 2 \rho R (T^{eq}
\delta_{ij} - T_{ij}) \nonumber \\
&=& -p \delta_{ij} + \mu [
\partial_i U_j +
\partial_j  U_i - \frac{2}{3} \partial_k U_k
 \delta_{ij}],
  \label{eq:stress-ns} \eey
which is exactly the Navier-Stokes stress-strain relationship. In
the Navier-Stokes equations, the stress becomes symmetric tensor
which is constructed through rational mechanical analysis. However,
in the new gas dynamic equations, the "stress" tensor is
automatically a symmetric tensor due to the commutable property of
random particle velocities, i.e., $(u_i - U_i) (u_j -U_j ) =(u_j -
U_j) (u_i -U_i ) $. In order words, the present
 formulation gives a microscopic interpretation of the origin of symmetric
 stress tensor. At the same time, the heat flux transported in $k-$direction for the
thermal energy $\rho R T_{ij}$ becomes
$$q_{kij}= \frac{\tau \rho R^2}{\mbox{Pr}}
(T_{kl}
\partial_l T_{ij} + T_{il}
\partial_l T_{jk} + T_{jl}
\partial_l T_{ki} ) .$$
These results are consistent with the early analysis for the
one-dimensional flow \cite{xu07}.

In summary, the generalized gas dynamic equations have been derived
based on the multiple stage BGK model. With the added Prandtl number
Pr and the introduction of dynamic viscosity in the determination of
relaxation parameter $\tau$, the gas dynamic equations derived in
this section are a closed system, which is a natural extension of
the Navier-Stokes equations. Theoretically, the new gas dynamic
equations cover a wider flow regime than that of the Navier-Stokes
equations. The new constitutive relationship now has a microscopic
physical basis. The viscous term involves only first-order
derivatives of the flow variables. The ability to treat
non-equilibrium flow problems without evaluating higher than first
order derivatives would prove very advantages numerically. The
method should be relatively insensitive to irregularity of the grid,
the straight-forward computing using Discontinuous Galerkin
framework, and the easy implementation of boundary conditions. Due
to the low order equations, the less communication between the
computational cells (small stencil) makes its numerical scheme more
efficient to implement on parallel computing architecture.

\section{Solutions of the generalized  gas dynamic equations}

In this section, we are going to present two flow problems in the
near continuum flow regime for the validation of the generalized gas
dynamic equations derived in the last section. Both test problems
are microchannel flows, but with different flow speed and
non-equilibrium properties.

\subsection{Analytic solution of force-driven Poiseuille flow in near
continuum flow regime}

In this subsection, we will apply the generalized gas dynamic
equations to rarefied gas flows, such as the case of force driven
Poiseuille flow between two parallel plates. Both Direct Simulation
Monte Carlo (DSMC) and kinetic theory have shown that even with a
small Kn, the pressure and temperature profiles in this flow exhibit
a different qualitative behavior from those predicted by the
Navier-Stokes equations
\cite{ref:DSMC_Poi,ref:BGK_Poi,ref:Tij_Poi,ref:BGK_Aoki,ref:Zheng_JSP2002,ref:Xu_JFM2004}.
Therefore, this flow can serve as a good test problem for any
extended hydrodynamic equations intended for non-continuum flow
computation.

In our test, the two walls of the channel locate at $y=\pm H/2$,
respectively, and the force $\bm{a}=(a,0,0)$ is along the $x$
direction. The flow is assumed to be unidirectional, i.e,
$\partial_x\phi=0$ for any variable $\phi$, and the velocity has
only an $x$-component in the laminar and stationary case, i.e.,
$\bm{U}_m=(U,0,0)$. Under such conditions, the generalized gas
dynamic equations (\ref{eq:con-r})-(\ref{eq:ene-r}) reduce to
\bem
2\D{P_{xy}}{y}-\rho a=0, \eem
\bem
 \D{}{y}(2P_{yy}-P)=0,
\eem
\bem
 \frac{1}{\PP}\D{}{y}\left(2\tau P_{xy}\D{\theta_{xy}}{y}+ \tau
P_{yy}\D{\theta_{xx}}{y}\right)=\frac{2(P_{xx}-P)}{\tau}+4P_{xy}\D{U}{y},
\eem
\bem
 \frac{3}{\PP}\D{}{y}\left(\tau
P_{yy}\D{\theta_{yy}}{y}\right)=\frac{2(P_{yy}-P)}{\tau}, \eem
\bem \frac{1}{\PP}\D{}{y}\left(\tau
P_{yy}\D{\theta_{zz}}{y}\right)=\frac{2(P_{zz}-P)}{\tau}, \eem
and
\bem
 \frac{1}{\PP}\D{}{y}\left(2\tau P_{yy}\D{\theta_{xy}}{y}+
\tau
P_{xy}\D{\theta_{yy}}{y}\right)=\frac{2P_{xy}}{\tau}+(2P_{yy}-P)\D{U}{y},
\eem
where $P_{ij}=\rho \theta_{ij}$ is the pressure tensor, and $P=\rho
\theta$ is the pressure, with $\theta_{ij}=RT_{ij}$ and $\theta=R
T^{eq}$.

It is difficult to obtain an analytical solution from the above
nonlinear system. Here we will try to find an approximate solution
using a perturbation method similar to that used in Ref.
\cite{ref:Tij_Poi}. To this end, we first introduce the following
dimensionless variables:
$$
\hat{y}=\frac{y}{H},\quad \hat{\rho}=\frac{\rho}{\rho_0},\quad
\hat{P}_{ij}=\frac{P_{ij}}{P_0},\quad
\hat{\theta_{ij}}=\frac{\theta_{ij}}{\theta_0}, \quad
\hat{U}=\frac{U}{\sqrt{\theta_0}},\quad
\hat{\tau}=\frac{\tau}{\tau_0},
$$
where the variables with subscript $0$ represent the corresponding
reference quantities. With these dimensionless variables, the system
can be rewritten as
\begin{equation}
\label{eq:pxy} \D{\hat{P}_{xy}}{\hat{y}}-\epsilon\hat{\rho}=0,
\end{equation}
\begin{equation}
\label{eq:C}
2\hat{P}_{yy}-\hat{P}=C,
\end{equation}

\begin{equation}
\label{eq:pxx}
\D{}{\hat{y}}\left(2\hat{\tau}\hat{P}_{xy}\D{\theta_{xy}}{\hat{y}}+
\hat{\tau}\hat{P}_{yy}\D{\hat{\theta}_{xx}}{\hat{y}}\right)=
2\mbox{Pr}\delta^2\frac{(\hat{P}_{xx}-P)}{\tau}+4\mbox{Pr}\delta\hat{P}_{xy}\D{\hat{U}}{\hat{y}},
\end{equation}

\begin{equation}
\label{eq:pyy}
\D{}{\hat{y}}\left(\hat{\tau}\hat{P}_{yy}\D{\hat{\theta}_{yy}}{\hat{y}}\right)=
2\mbox{Pr}\delta^2\frac{(\hat{P}_{yy}-\hat{P})}{\tau},
\end{equation}

\begin{equation}
\label{eq:pzz}
\D{}{\hat{y}}\left(\hat{\tau}\hat{P}_{yy}\D{\hat{\theta}_{zz}}{\hat{y}}\right)=
2\mbox{Pr}\delta^2\frac{(\hat{P}_{zz}-\hat{P})}{\tau},
\end{equation}
and
\begin{equation}
\label{eq:U}
\D{}{\hat{y}}\left(2\hat{\tau}\hat{P}_{yy}\D{\hat{\theta}_{xy}}{\hat{y}}+
\hat{\tau}\hat{P}_{xy}\D{\hat{\theta}_{yy}}{\hat{y}}\right)=
2\mbox{Pr}\delta^2\frac{\hat{P}_{xy}}{\tau}+C\mbox{Pr}\delta\D{\hat{U}}{\hat{y}},
\end{equation}
where $C$ is a constant, and $\epsilon$ and $\delta$ are two
dimensionless parameters given by
$$
\epsilon=\frac{a}{2\theta_0},\quad
\delta=\frac{H}{\tau_0\sqrt{\theta_0}}=\sqrt{\frac{\pi}{2}}\frac{1}{\mbox{Kn}}.
$$

Now we assume that the force acceleration $a$ is small such that
$\epsilon\ll 1$, then we can expand the dimensionless flow
quantities in powers of $\epsilon$ (we will omit the hat for
simplicity hereafter):
$$
\rho=1+\epsilon^2\rho^{(2)}+O(\epsilon^4),\quad U=\epsilon
U^{(1)}+O(\epsilon^3),\quad C=1+\epsilon^2 C^{(2)},
$$
$$
P_{xy}=\epsilon P_{xy}^{(1)}+O(\epsilon^3),\quad
\theta_{xy}=\epsilon\theta_{xy}^{(1)}+O(\epsilon^3),
$$
$$
P_{ii}=1+\epsilon^2 P_{ii}^{(2)}+O(\epsilon^4),\quad
\theta_{ii}=1+\epsilon^2\theta_{ii}^{(2)}+O(\epsilon^4),\quad
{\mbox{for}} \quad  i=x,y,z,
$$
and
$$
P=1+\epsilon^2 P^{(2)}+O(\epsilon^4),\quad
\theta=1+\epsilon^2\theta^{(2)}+O(\epsilon^4),\quad
\tau=1+\epsilon^2\tau^{(2)}+O(\epsilon^4) .
$$
The odd or even properties of the variables as functions of
$\epsilon$ are based on their symmetric properties in terms of the
acceleration $a$. In general, the odd(even) velocity moments of the
distribution function $f$ are also odd (even) functions of
$\epsilon$. Furthermore, from the definition $P_{ij}=\rho
\theta_{ij}$ we can obtain some useful relation which will be used
later:
\begin{equation}
\label{eq:rho2}
\rho^{(2)}=P^{(2)}-\theta^{(2)}=P_{ii}^{(2)}-\theta_{ii}^{(2)},
\quad P_{xy}^{(1)}=\theta_{xy}^{(1)}, \quad \mbox{for} \quad i=x, y,
z.
\end{equation}

Substituting the above expansions into the nondimensional system, we
can obtain the first order differential equation in $\epsilon$. Eqs.
(\ref{eq:pxy}) and ({\ref{eq:U}}) go to
\bem \D{P_{xy}^{(1)}}{y}-1=0, \eem
\bem
\mbox{Pr}\delta\D{U^{(1)}}{y}=\D{^2\theta_{xy}^{(1)}}{y^2}-2\mbox{Pr}\delta^2P_{xy}^{(1)},
\eem
 which give
\begin{equation}
\label{eq:pxy1} P_{xy}^{(1)}=\theta_{xy}^{(1)}=y,
\end{equation}
\begin{equation}
\label{eq:U1} U^{(1)}=-\delta y^2+C',
\end{equation} where $C'$
is a constant. Here, we have made use of the symmetry property of
the velocity profile about $y=0$.

From Eqs. (\ref{eq:C}) (\ref{eq:pxx}), (\ref{eq:pyy}), and
(\ref{eq:pzz}), we can obtain the second order equations in
$\epsilon$:
\begin{equation}
\label{eq:C2} 2P_{yy}^{(2)}-P^{(2)}=C^{(2)},
\end{equation}

\begin{equation}
\label{eq:pxx2}
\D{^2\theta_{xx}^{(2)}}{y^2}=2\mbox{Pr}\delta^2\left(\theta_{xx}^{(2)}-\theta^{(2)}\right)-1-8\mbox{Pr}\delta^2y^2,
\end{equation}

\begin{equation}
\label{eq:pyy2}
3\D{^2\theta_{yy}^{(2)}}{y^2}=2\mbox{Pr}\delta^2\left(\theta_{yy}^{(2)}-\theta^{(2)}\right)
\end{equation}
and
\begin{equation}
\label{eq:pzz2}
\D{^2\theta_{zz}^{(2)}}{y^2}=2\mbox{Pr}\delta^2\left(\theta_{zz}^{(2)}-\theta^{(2)}\right)
\end{equation}
where the results up to the first order Eqs. (\ref{eq:pxy1}) and
(\ref{eq:U1}) have been used. Summing up the above three equations
we have
\begin{equation}
\label{eq:p2}
\D{^2}{y^2}\left(3\theta^{(2)}+2\theta_{yy}^{(2)}\right)=-1-8\mbox{Pr}\delta^2y^2.
\end{equation}
From Eqs. (\ref{eq:pyy2}) and (\ref{eq:p2}), we can get
\bem
 \theta^{(2)}=-\frac{2}{3}A\cosh\left(\frac{\sqrt{10}}{3}K
y\right) -\frac{2K^2}{15}y^4 +
\frac{43}{50}y^2+\frac{231}{125K^2}+B, \eem
\bem
 \theta_{yy}^{(2)}=A\cosh\left(\frac{\sqrt{10}}{3}K y\right)
-\frac{2K^2}{15}y^4 -\frac{77}{50}y^2-\frac{693}{250K^2}+B.
\eem
Since $\theta^{(2)}$ is obtained, the other two temperatures
$\theta_{xx}^{(2)}$ and $\theta_{zz}^{(2)}$ from Eqs.
(\ref{eq:pxx2}) and (\ref{eq:pzz2}) can be constructed,
respectively.

 Now let's find the second order pressure $P^{(2)}$. From Eqs.
(\ref{eq:rho2}) and (\ref{eq:C2}), we can obtain
\bem
 P^{(2)}=C^{(2)}+2(\theta^{(2)}-\theta_{yy}^{(2)})
=C^{(2)}-\frac{10A}{3}\cosh\left(\frac{\sqrt{10}}{3}K y\right) +
\frac{24}{5}y^2+\frac{231}{25K^2}. \eem
Then, $\rho^{(2)}$ can be determined from (\ref{eq:rho2}) as
\bem \rho^{(2)}=P^{(2)}-\theta^{(2)})
=C^{(2)}-\frac{8A}{3}\cosh\left(\frac{\sqrt{10}}{3}K y\right) +
\frac{2K^2}{15}y^4+\frac{197}{50}y^2+\frac{924}{125K^2}-B.
\eem

With the above results, we finally get the approximate solutions of
the problem:
\begin{equation}
\label{eq:uABC}
\frac{U}{\sqrt{RT_0}}=-\epsilon\delta \left(\frac{y}{H}\right)^2+U_s,
\end{equation}
\begin{equation}
\label{eq:pABC}
\frac{P}{P_0}=1+\epsilon^2 P^{(2)},
\end{equation}
\begin{equation}
\label{eq:TABC}
\frac{T}{T_0}=1+\epsilon^2 \theta^{(2)}.
\end{equation}

With these results, we are  able to make some discussions on the
velocity, pressure, and temperature profiles for the force-driven
Poiseuille flow problem. First, it is clear that the velocity
profile is parabolic. But for the pressure and temperature, their
profiles may be complicated due to the presence of the hyperbolic
cosine function.

It is interesting to compare the approximate solutions of the
present gas dynamic equations to those of the Navier-Stokes-Fourier
(NSF) equations. For the same flow problem, the NSF equations reduce
to
\begin{equation}
\label{eq:NSu}
\D{}{y}\left(\mu\D{U}{y}\right)+\rho a =0,
\end{equation}
\begin{equation}
\label{eq:NSp}
\D{P}{y} = 0,
\end{equation}
\begin{equation}
\label{eq:NST}
\D{}{y}\left(k\D{U}{y}\right)+\mu\left(\D{U}{y}\right)^2 =0.
\end{equation}
Using the similar perturbation method, we can get the following
approximate solutions for the NSF equations:
\begin{equation}
\label{eq:uNSa}
\frac{U}{\sqrt{RT_0}}=-\epsilon\delta \left(\frac{y}{H}\right)^2+U_s,\quad P=\mbox{const},\quad
\frac{T}{T_0}=1+\epsilon^2 \theta_{nsf}^{(2)},
\end{equation}
where $\theta_{nsf}^{(2)}=-(2/15)\mbox{Pr}\delta^2 \hat{y}^4+D$, and
$D$ is a constant. It is shown that the velocity profile of the NSF
equations is also parabolic, which is the same as the that of the
generalized hydrodynamic equations. However, the NSF equations give
a constant pressure, which is qualitatively different from the
predictions of the new proposed model. For the temperature, it is
evident that $\theta_{nsf}^{(2)}$ is only part of $\theta^{(2)}$,
and has one local maximum at $y=0$. On the other hand, the presence
of terms of $\cosh(y)$ and $y^2$ in $\theta^{(2)}$ of the new model
makes it possible to have a local minimum at $y=0$, due to
$[\theta^{(2)}]'_{y=0}=0$ and
$[\theta^{(2)}]''_{y=0}=-K^2(100A+216)/135+1.72>0$ as
$A<2.322K^{-2}-2.16$.

To see this more clearly, we present the pressure and temperature
profiles for the case of Kn=0.1 which was studied extensively using
DSMC method \cite{ref:Zheng_JSP2002}. The constants $A$ and $B$ in
$\theta$ and $P$ of the present hydrodynamic model are obtained by
enforcing the values of $\theta$ at $y=0$ and $-H/2$ to be the same
as the DSMC data, and then $C$ is obtained by enforcing $P(y=0)$ to
be the DSMC value. The NSF solutions are obtained by enforcing their
pressure and temperature values to be identical to the DSMC data at
$y=0$. In Fig.\ref{fig:2}, the reduced temperature and pressure
variations, $\theta^{(2)}=(T/T_0-1)/\epsilon^2$ and
$P^{(2)}=(P/P_0-1)/\epsilon^2$, are shown for both the NSF equations
and the present model.  It is seen that the temperature and pressure
profiles of the present model are in qualitatively agreement with
the DSMC data. For example, the temperature takes a bimodal shape
and exhibits a local minimum at $y=0$, and the pressure has two
local maximums near the two walls. These critical flow behaviors are
absent in the profiles of the NSF equations. For example, the
temperature minimum only appears on the super-Burnett order if the
traditional BGK collision model is used to construct the gas dynamic
equations \cite{chapman,xu-super}. These observations demonstrate
the fundamental difference between the present hydrodynamic model
and the NSF equations.

\subsection{Numerical solution for Couette Flows}

For the new gas dynamic equations, a corresponding gas-kinetic
scheme can be developed \cite{xu07}. In order to further test the
validity of the new governing equations, we will study the planar
Couette flow here \cite{alex}. The height of the Couette system is
$h_0 = 50 nm$. The wall's temperatures are fixed at $T_0 = 273K$ and
at equilibrium, the pressure of the gas is $1$ atm. Given that the
walls distance is less than a mean-free path and the relative wall
speed is high, the gas system will be strongly out of equilibrium.
Specifically, the velocity distribution for the particles is
non-Maxwellian. Same as the case in \cite{alex}, the temperature
along the channel is defined as $T_x$ and the one perpendicular to
walls is $T_z$. With the fixed Knudsen number Kn$=1.25$, and various
wall velocities, such as the Mach number $M=0.5,1.0,$ and $1.5$, the
simulation results from the new gas dynamic equations are shown in
Fig.\ref{fig:3}, where the DSMC solutions are also included
\cite{alex}. Clearly, at high Knudsen number and Mach numbers, the
temperature is anisotropic. The new gas dynamic equations basically
are capable in capturing this kind of non-equilibrium flow
phenomena. Also, it is fully necessary for any gas dynamic equations
to consider the temperature as a tensor for the non-equilibrium
system.

\section{Conclusion}

Based on the multiple stage particle collision BGK model and the
Gaussian distribution function as the middle state, the generalized
gas dynamic equations have been derived. Since the gas temperature
basically represents the molecular random motion,  the direct
extension of the temperature concept from a scalar to a second-order
symmetric tensor $T_{ij}$ is physically reasonable. In the
non-equilibrium flow regime, the randomness of the particle
distribution indeed depends on the spatial orientation. The new gas
dynamic equations have the same structure as the Navier-Stokes
equatons, but the NS constitutive relationship,
$$\sigma_{ij} = -\rho R T^{eq} \delta_{ij} + \mu (\partial_i U_j +
\partial_j U_i - \frac{2}{3} \partial_k U_k \delta_{ij} ) $$ is
replaced by
$$\sigma_{ij} = -\rho R T_{ij} + \rho R (T^{eq} \delta_{ij} -
T_{ij}  ) .$$
 At the same time, the heat flux in the
$k-$direction for the transport of thermal energy $\rho R T_{ij}$
becomes
$$q_{kij}= \frac{\tau \rho R^2}{\mbox{Pr}} (T_{kl}
\partial_l T_{ij} + T_{il}
\partial_l T_{jk} + T_{jl}
\partial_l T_{ki} ) .$$
In the continuum flow regime, the generalized constitutive
relationship and the heat flux term go back to the corresponding
Navier-Stokes formulations. The new gas dynamic equations can be
regarded as a regularization of Levermore's 10 moment closure
\cite{levermore}. The gas dynamic equations have a wider applicable
flow regime than that of the NS equations. They capture the time
evolution of the anisotropic non-equilibrium flow variables, as
demonstrated in our examples.

The traditional temperature concept is coming from thermodynamics,
where there is no anisotropic particle random motion in space.
However, for the non-equilibrium flow transport, due to the
inadequate of particle collision the random molecule motion can
become easily anisotropic. To directly consider the temperature as a
tensor rather than a scalar is a reasonable description for the
non-equilibrium flow. For a dilute gas, due to the lack of long
range particle interaction, the randomness particle motion is the
only source for the dissipation in the system. Under the new
definition of the temperature, all dissipative effects in a dilute
gas system, such as the viscosity and heat conduction, can be
unified under the same concept $T_{ij}$. The current gas dynamic
equations can be useful in the study of microflows. The further
development of the gas dynamic equations based on the kinetic model
and its scheme in \cite{xu-eswar}, which are valid for the
compressible shock waves, will be conducted in the near future.

\section*{Acknowledgement}
K. Xu would like to thank Prof. C. Groth and Mr. J.G. McDonald for
helpful discussion, and Dr. H.W. Liu for his help in the numerical
testing of Couette flow. This research was supported by Hong Kong
Research Grant Council 621005. Z. Guo acknowledges the support of
the National Natural Science Foundation of China (50606012).

\section*{Appendix: Gas Dynamic Equations in 2D Space}

As a special application, we consider 2D case, where there are  $7$
unknowns $(\rho, U, V, T_{xx}, T_{xy}, T_{yy}, T_{zz}) $. The
generalized gas dynamic equations for those unknowns are
\beq \frac{\partial \bf{U}}{\partial t} + \frac{\partial
\bf{F}}{\partial x} + \frac{\partial \bf{G}}{\partial y} =
\frac{\partial {\bf E_v} }{\partial x} + \frac{\partial {\bf
F_v}}{\partial y} + {\bf S}, \eeq
where $\bf U$ is the vector macroscopic flow variables, and $\bf F$
and $\bf G$ are $x$ and $y$ direction flux vectors given by
$$
{\bf U} = \pmatrix{ \rho \cr
            \rho U \cr
           \rho V   \cr
           \rho (U^2 + R T_{xx})  \cr
            \rho (UV + RT_{xy} )\cr
           \rho (V^2 + RT_{yy} ) \cr
           \rho R T_{zz}  },
{\bf F} =\pmatrix{ \rho U \cr
            \rho (U^2 + R T_{xx} ) \cr
           \rho (UV + R T_{xy} )   \cr
           \rho (U^3 + 3 U R T_{xx})  \cr
            \rho (U^2V + 2U RT_{xy} + V RT_{xx} )\cr
           \rho (UV^2 + U RT_{yy} + 2V RT_{xy}) \cr
           \rho U R T_{zz}  },
$$

$$
{\bf G} =\pmatrix{ \rho V \cr
            \rho (U V + R T_{xy} ) \cr
           \rho (V^2 + R T_{yy} )   \cr
           \rho (U^2 V + V R T_{xx} + 2U RT_{xy})  \cr
            \rho (UV^2 + U RT_{yy} + 2 V RT_{xy} )\cr
           \rho (V^3 + 3V RT_{yy}) \cr
           \rho V R T_{zz}  },
$$
$$
{\bf E_v} =\pmatrix{ 0 \cr
            \rho R (T^{eq}\delta_{xx} - T_{xx} ) \cr
           \rho R (T^{eq}\delta_{xy} - T_{xy} )   \cr
           3 \rho R [ U (T^{eq}\delta_{xx} - T_{xx}) + \tau R (T_{xx} \partial_x T_{xx} + T_{xy} \partial_y T_{xx} )] \cr
            \rho R [ V (T^{eq}\delta_{xx} - T_{xx} ) + 2 U (T^{eq}\delta_{xy} - T_{xy} + \tau R (T_{xy} \partial_x T_{xx} + T_{yy} \partial_y T_{xx}
            + 2 T_{xx}\partial_x T_{xy}+ 2 T_{xy} \partial_y T_{xy}) )]\cr
           \rho R [U (T^{eq}\delta_{yy} - T_{yy})+ 2 V (T^{eq}\delta_{xy} -T_{xy} ) + \tau R (2 T_{xy}\partial_x T_{xy} + 2 T_{yy} \partial_y T_{xy}
           +T_{xx} \partial_x T_{yy} + T_{xy} \partial_y T_{yy} )   ] \cr
           \rho R [U(T^{eq}\delta_{zz} - T_{zz} ) + \tau R (T_{xx}\partial_x T_{zz} + T_{xy} \partial_y T_{zz} ) ]
           },
$$

$$
{\bf F_v} =\pmatrix{ 0 \cr
            \rho R (T^{eq}\delta_{xy} - T_{xy} ) \cr
           \rho R (T^{eq}\delta_{yy} - T_{yy} )   \cr
\rho R [V (T^{eq}\delta_{xx} - T_{xx})+ 2 U (T^{eq}\delta_{xy}
-T_{xy} ) + \tau  R(2 T_{xy}\partial_y T_{xy} + 2 T_{xx} \partial_x
T_{xy}
           +T_{yy} \partial_y T_{xx} + T_{xy} \partial_x T_{xx} )   ] \cr
        \rho R [ U (T^{eq}\delta_{yy} - T_{yy} ) + 2 V (T^{eq}\delta_{xy} - T_{xy} + \tau R(T_{xy} \partial_y T_{yy} + T_{xx} \partial_x T_{yy}
            + 2 T_{yy}\partial_y T_{xy}+ 2 T_{xy} \partial_x T_{xy}) )]\cr
           3 \rho R [ V (T^{eq}\delta_{yy} - T_{yy}) + \tau R (T_{yy} \partial_y T_{yy} + T_{xy} \partial_x T_{yy} )] \cr
           \rho R [V(T^{eq}\delta_{zz} - T_{zz} ) + \tau R (T_{yy}\partial_y T_{zz} + T_{xy} \partial_x T_{zz} ) ]  },
$$
and the source term is
$$
{\bf S} =\pmatrix{ 0 \cr
             0 \cr
             0  \cr
           2\rho R ( T^{eq}\delta_{xx} - T_{xx})/\tau  \cr
            2\rho R (T^{eq}\delta_{xy} -T_{xy} )/\tau \cr
           2\rho R (T^{eq}\delta_{yy} - T_{yy} )/\tau \cr
           2\rho R (T^{eq}\delta_{zz} -T_{zz} ) /\tau }.
$$
In the above equations, the equilibrium temperature
$$ T^{eq} = T^{eq}\delta_{xx} = T^{eq}\delta_{yy} = T^{eq}\delta_{zz} = \frac{1}{3} \mbox{tr}
(T_{ij}) ,$$ and
$$T^{eq}\delta_{xy} = T^{eq}\delta_{yx} = 0 .$$

\begin{figure}
\begin{center}
\scalebox{0.5}{\includegraphics{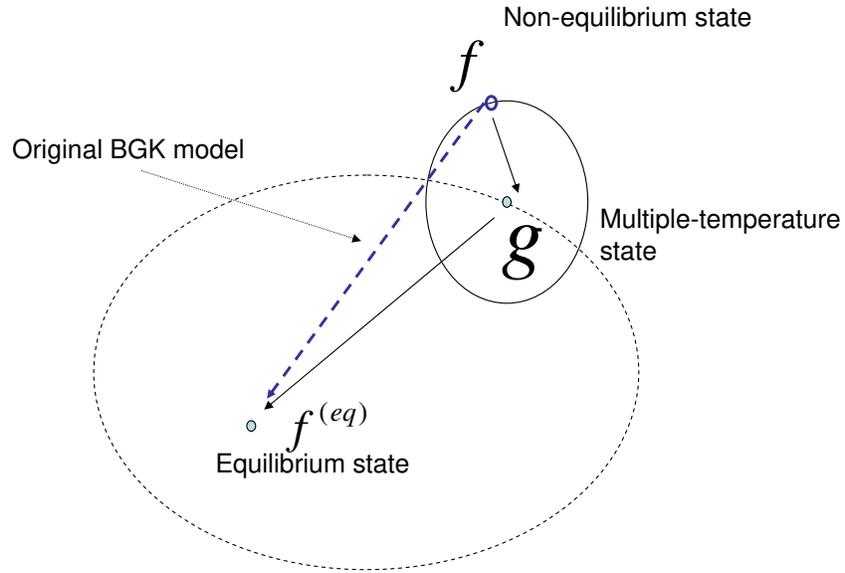}}
\caption{Schematic representation of particle collision relaxation
processes.}
\label{fig:1}
\end{center}
\end{figure}

\begin{figure}
\begin{center}
\scalebox{0.5}{\includegraphics{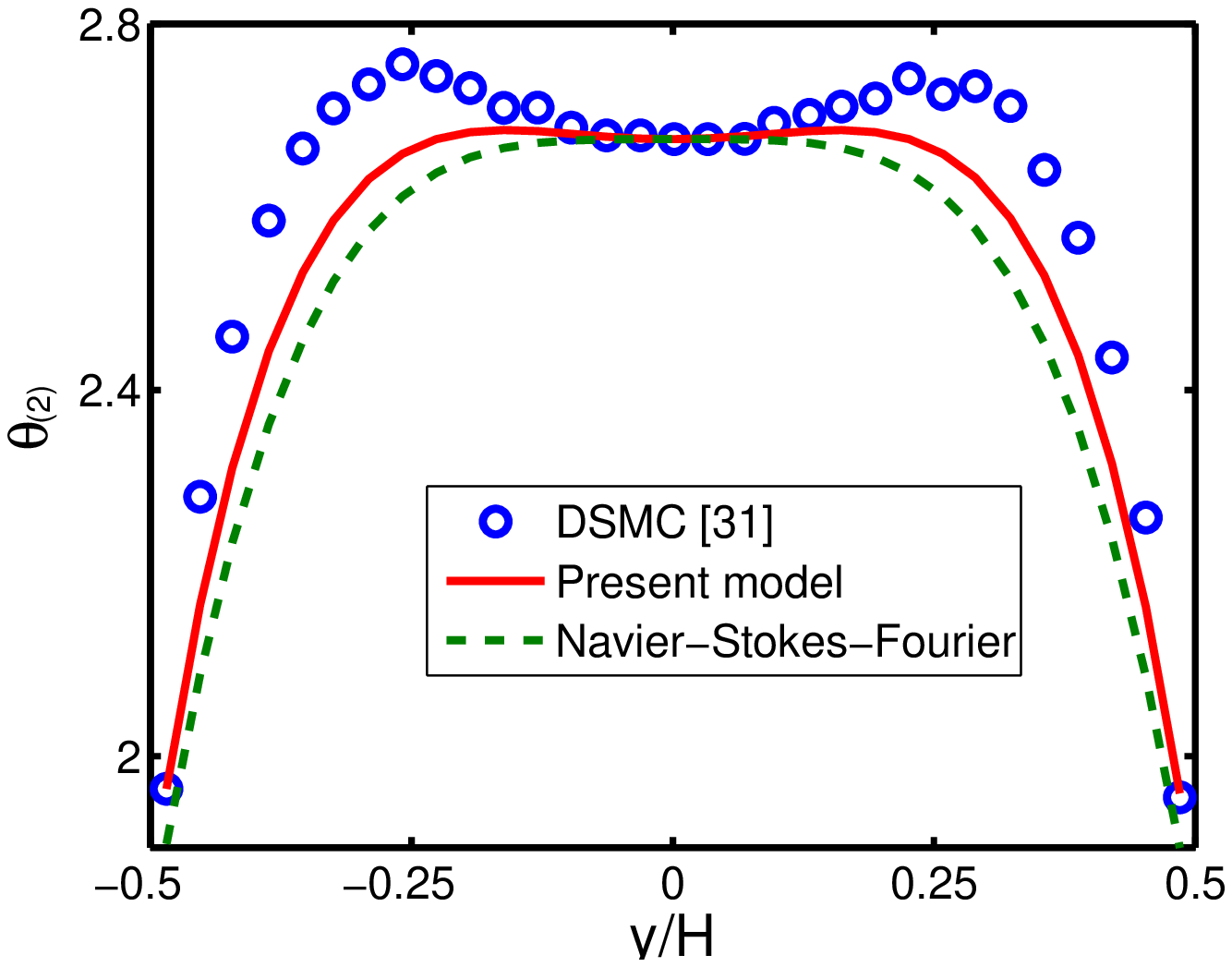}}
\scalebox{0.5}{\includegraphics{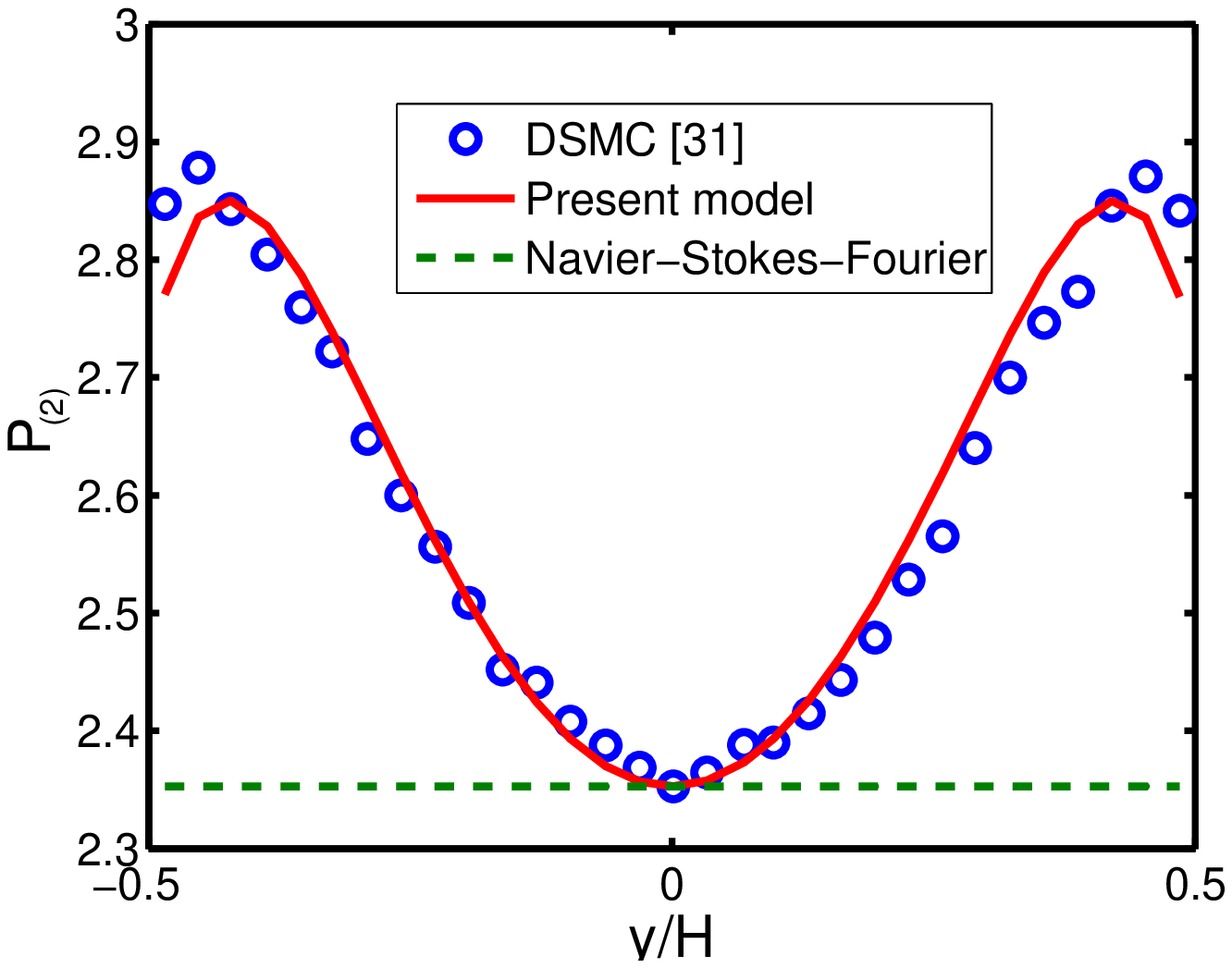}} \caption{Reduced
temperature (up) and pressure (down) variations at $Kn=0.1$, where
the results from the DSMC \cite{ref:Zheng_JSP2002}, the current
equations, and the Navier-Stokes equations are presented.}
\label{fig:2}
\end{center}
\end{figure}

\begin{figure}
\begin{center}
\scalebox{0.5}{\includegraphics{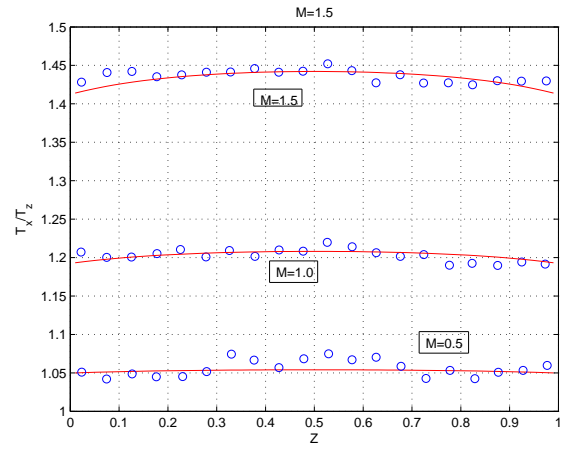}}
\caption{Ratio of the temperature components $T_x/T_z$ vs the
vertical position $Z=z/h_0$ in planar Couette flow for Knudsen
number Kn=1.25, and Mach numbers $M=$ $0.5, 1.0,$ and $1.5$. The
solid lines are the solutions of the new gas dynamic equations and
the circles are the DSMC solutions \cite{alex}.} \label{fig:3}
\end{center}
\end{figure}

\end{document}